\documentclass{emulateapj}
\usepackage{apjfonts}

\begin{document}

\title{Disentangling hadronic and leptonic cascade scenarios from the very-high-energy gamma-ray emission of distant hard-spectrum blazars}

\author{Hajime Takami\altaffilmark{1}, Kohta Murase\altaffilmark{2}, Charles D. Dermer\altaffilmark{3}}

\altaffiltext{1}{JSPS Research Fellow --- Institute of Particle and Nuclear Studies, KEK, 1-1, Oho, Tsukuba 305-0801, Japan. e-mail: takami@post.kek.jp}
\altaffiltext{2}{Hubble Fellow --- Institute for Advanced Study, 1 Einstein Dr. Princeton, NJ 08540, USA. e-mail: murase@ias.edu}
\altaffiltext{3}{Code 7653, Space Science Division, U.S. Naval Research Laboratory, Washington, DC 20375, USA. e-mail: charles.dermer@nrl.navy.mil}

\begin{abstract}
Recent data from the \emph{Fermi} Large Area Telescope have revealed about a dozen distant hard-spectrum blazars that have very-high-energy (VHE; $\gtrsim 100$ GeV) photons associated with them, but most of them have not yet been detected by imaging atmospheric Cherenkov telescopes. Most of these high-energy gamma-ray spectra, like those of other extreme high-frequency peaked BL Lac objects, can be well explained either by gamma rays emitted at the source or by cascades induced by ultra-high-energy cosmic rays, as we show specifically for KUV 00311$-$1938. We consider the prospects for detection of the VHE sources by the planned Cherenkov Telescope Array (CTA) and show how it can distinguish the two scenarios by measuring the integrated flux above $\sim 500$ GeV (depending on source redshift) for several luminous sources with $z \lesssim 1$ in the sample. Strong evidence for the origin of ultra-high-energy cosmic rays could be obtained from VHE observations with CTA. Depending on redshift, if the often quoted redshift of KUV 00311-1938 ($z = 0.61$) is believed, preliminary H.E.S.S. data favor cascades induced by ultra-high-energy cosmic rays. Accurate redshift measurements of hard-spectrum blazars are essential for this study. 
\end{abstract}


\section{Introduction} \label{sec:introduction}

Gamma-ray astronomy has rapidly advanced in recent years due to results from the Large Area Telescope (LAT) on board the \emph{Fermi} Gamma-ray Space Telescope, which is sensitive to high-energy gamma rays from $20$ MeV to $\gtrsim 300$ GeV \citep{Atwood2009ApJ697p1071}, and from Imaging Atmospheric Cherenkov Telescopes (IACTs), which detect  very-high-energy (VHE; $> 100$ GeV) gamma rays. The survey capability and wide energy range of the LAT also provides useful information on transient activities of known gamma-ray sources and potential VHE emitters.

In a recent analysis of LAT data, searches for sources of VHE photons found thirteen promising candidates of VHE gamma-ray emission at high ($z > 0.5$) redshifts \citep{Neronov2012arXiv1207.1962}. The detections of KUV 00311$-$1938, PKS 0426$-$380, and 4C +55.17 was first reported in the LAT Bright Active Galactic Nuclei Source List \citep[]{Abdo2009ApJ700p597}. All these sources except PKS 2142$-$75 are found in the First LAT AGN Catalog \citep[]{Abdo2010ApJ715p429}, and all appear in the Second LAT AGN Catalog \citep{Ackermann2011ApJ743p171}.

Depending on source redshift, VHE gamma rays above a characteristic energy $E_c$ are absorbed by photons of the extragalactic background light (EBL) and the cosmic microwave background (CMB) during propagation through intergalactic space, and create electron-positron pairs \citep[e.g.,][]{Nikishov1962JETP14p393,Gould1966PRL16p252,Stecker1992ApJ390L49}. The value of $E_c$, defined by the condition $\tau_{\gamma\gamma}(E_c,z) = 1$ that the $\gamma\gamma$ opacity is unity, is lower for higher redshift sources. For recent EBL models \citep[e.g.,][]{Kneiske2004AA413p807,Finke2010ApJ712p238,Dominguez2011MNRAS410p2556,Inoue2012arXiv1212.1683}, $E_c\sim 200$ GeV for $z = 0.5$ and $E_c\sim 100$ GeV for $z = 1$. The detection of VHE gamma rays from distant sources therefore provides useful information on the formation of stars that make the EBL.

The secondary electrons and positrons (hereafter called electrons unless stated otherwise) produced by $\gamma\gamma$ pair creation generate gamma rays by Compton-scattering target photons of the EBL and CMB, initiating an electromagnetic cascade if the produced gamma rays are subsequently absorbed. The resultant secondary photons also contribute to the gamma-ray flux. Their fluxes are also affected by intergalactic magnetic fields (IGMFs) due to magnetic deflection of the  parent electrons and produce, depending in detail on the strength of the IGMF and its coherence length, a gamma-ray halo \citep[e.g.,][]{Aharonian1994ApJ423L5,Elyiv2009PRD80p023010} and a delayed gamma-ray component \citep[pair echo; e.g.,][]{Plaga1995Nature374p430,Murase2008ApJ686L67}. Combined  LAT and IACT fluxes of extreme high-frequency peaked BL Lac objects (HBLs) have recently been used to constrain the strength of IGMFs  \citep[e.g.,][]{Neronov2010Science328p73,Dolag2011ApJ727L4,Dermer2011ApJ733L21}.

Ultra-high-energy (UHE; $\gtrsim 10^{18}$ eV) cosmic rays (CRs) are also sources of electromagnetic cascades. Interactions of UHECRs with CMB and EBL photons produce UHE electrons and photons through Bethe-Heitler pair creation and photomeson production \citep[e.g.,][]{Wdowczyk1972JPhA5p1419,Yoshida:1993pt,Lee1998PhRvD58p043004}. If cosmic magnetic fields are weak enough, $\ll 10^{-10}$ G, to allow protons to propagate almost rectilinearly, which always holds in the voids for the magnetic fields considered here, though not always in the structured region \citep[][]{Murase2012ApJ749p63}, high-energy gamma rays are produced from electromagnetic cascades induced by these particles. This process has been proposed as a solution to explain the observed spectra of extreme HBLs \citep[e.g.,][]{Essey2010APh33p81,Essey2010PRL104p141102,Essey2011ApJ731p51,Murase2012ApJ749p63}.

In this Letter, we show that both gamma-ray and UHECR-induced cascade scenarios can,  in the case of weak IGMFs, reproduce the spectral energy distributions (SEDs) of most of the VHE candidates listed in the  \citet{Neronov2012arXiv1207.1962} LAT sample based on $\sim 4$ years of data. It then becomes important to distinguish these two scenarios observationally. If UHECRs are required to reproduce future VHE observations, this would provide strong evidence for UHECR origin in these sources and help solve a major  problem in high-energy astrophysics. In the cases of nearby extreme HBLs such as 1ES 0229+200, we have previously shown that gamma-ray spectra above $\sim 25$ TeV give a hint for the distinction because the hadronic scenario predicts harder spectra above $10$ TeV than the gamma-ray-induced cases \citep{Murase2012ApJ749p63}. Here we demonstrate whether such a distinction can be obtained for VHE candidates in future observations. Throughout this letter, a $\Lambda$CDM cosmology with $\Omega_{\rm M} = 0.3$, $\Omega_{\Lambda} = 0.7$, and $H_0 = 71$ km s$^{-1}$ Mpc$^{-1}$ is assumed.

\section{Data Selection and Analysis} \label{sec:data}

We analyzed public \emph{Fermi} LAT Pass 7 data taken from 2008 August 4 to 2012 July 18, and derived spectra of the sources in the $>100$ GeV list following the recommended procedure and parameters described at the \emph{Fermi} Science Support Center (FSSC).\footnote{http://fermi.gsfc.nasa.gov/ssc/data/analysis/} We used the LAT standard analysis software, {\it ScienceTools} v9r27p1, with the instrument response functions P7SOURCE\_V6. The unbinned analysis was performed for {\it Source} class events with the zenith angle less than $100^{\circ}$ to reduce Earth-limb gamma-rays. The Region of Interest and the Source Region used to derive fluxes were chosen to be $8^\circ$ and $15^\circ$ in radius, respectively, which are large enough for analysis restricted to $\geq 10$ GeV energies. Four equal-width logarithmic bins per decade were used  to calculate the differential spectra of the sources. When gamma-ray data were fitted, we took sources listed in LAT 2-year Gamma Ray Source (2FGL) Catalog \citep{Nolan2012ApJS199p31} into account in the Source Region. We treated flux normalizations as free parameters and assumed the same spectral indices as in the catalog justified by the narrow energy bins. The resultant spectra are consistent with those of \citet{Neronov2012arXiv1207.1962}, where {\it UltraClean} events are adopted, and the 2FGL results. When no photon or just one photon was detected, or a TS smaller than 9 (corresponding to $\sim 3 \sigma$) was obtained for an energy bin, a 95\% confidence level upper limit was calculated by the Bayesian method provided in \emph{ScienceTools}. Light curves were calculated in the same manner in each time bin. 

\section{Gamma-ray-induced and UHECR-induced Cascades} \label{sec:confronting}

Electromagnetic cascades induced by VHE gamma rays emitted from sources are described by transport equations \citep[e.g.,][]{Lee1998PhRvD58p043004,Murase2012JCAP10p043}. We may use one-dimensional results when the observed flux is not smeared by the spatial extension due to IGMFs, and such IGMFs are required to explain VHE point sources that are observed. The interaction rates of pair creation and inverse Compton-scattering in the equations depend on the specific EBL model. Two models developed by \citet{Kneiske2004AA413p807} ({\it low IR} and {\it best fit}) are adopted for calculations and to demonstrate the uncertainty of EBL estimation.

The strength $B$ of the IGMF assumed in this Letter is clarified as follows. The flux of secondary gamma rays is suppressed if electrons in cascades are deflected by an angle larger than the jet opening angle $\theta_j$. The deflection angle of electrons by the IGMF is $\theta_{\rm IG} \sim {\rm min}[\ell_{\rm IC} / r_{\rm L},\sqrt{\lambda \ell_{\rm IC}}/r_{L}]$, where $\lambda$ is the coherence length and $\ell_{\rm IC} = 3 m_e^2 c^4 / 4 \sigma_T E_e u_{\rm CMB}$ is the inverse Compton energy-loss length of electrons with energy $E_e$ and Larmor radius $r_L = E_e / eB$. Here $u_{\rm CMB}$ is the energy density of the CMB at redshift $z$. The characteristic energy $E_{\gamma}$ of secondary gamma rays produced when electrons Compton-scatter CMB photons with mean energy $E_{\rm CMB}$ is $E_{\gamma} \sim (E_e / m_e)^2 E_{\rm CMB}$. The condition $\theta_{\rm IG} < \theta_j$ implies $B \lesssim 10^{-15} (E_{\gamma} / 10~{\rm GeV}) (\theta_j / 0.1)$ G. Practically, lower limits to $B$ are obtained by calculating how much the flux should be suppressed \citep{Murase2008ApJ686L67}. Conservatively assuming that the blazar is active for only a few years, gamma-ray observations of extreme HBLs  give $B \gtrsim 10^{-18}$ -- $10^{-20}$ G \citep[e.g.,][]{Dermer2011ApJ733L21,Takahashi2012ApJ744L7}. Practically, $B = 10^{-18}$ G is assumed for calculations, which corresponds to an IGMF strength between these values, because flux suppression due to the IGMF is negligible above $\sim 1$ -- $3$ GeV for $B \lesssim (1$ -- $3)\times{10}^{-15}$ G.

The intrinsic spectrum of gamma rays, representing the gamma-ray source term in the photon transport equation, is assumed to have a power-law shape $\propto E^{-s}$. The index $s$ is chosen to reproduce the \emph{Fermi} spectra in the range $1$ GeV $<E< E_c$. The source spectrum  is assumed to extend to a maximum energy $100$ TeV. The minimum energy is taken to be $10^{9.75}$ eV in the case of $s \geq 2$, while it is not set for $s < 2$, because the low-energy cutoff is unimportant for hard-spectrum sources. Note that $s<2$ corresponds to the case where a cascade component dominates in the high-to-VHE energy ranges. The requirement that the observed high-energy \emph{Fermi}-LAT spectra are reproduced is used to normalize the intrinsic spectra of the sources.

In the case of UHECR-induced cascades, the source terms are replaced by the deposition rates of electrons and gamma rays at each point during UHECR propagation.  The deposition rates are calculated based on one-dimensional propagation of UHE protons, following calculations in \cite{Murase2012ApJ749p63}. The injection spectrum of protons with energy $E_p$ is assumed to be a power-law with an exponential cutoff of the form $E_p^{-p} \exp ( - E_p / E_{\rm p,c} )$. We take $p = 2.6$, $E_{\rm p,c} = 10^{19}$ eV, and the minimum energy of $10^{18}$ eV. The resultant cascade spectra are insensitive to the value of $p$, though the normalization is affected by $p$ and the UHECR energy-loss length due to Bethe-Heitler pair creation becomes shortest at $E_p \sim 10^{19}$ eV for $z \ll 1$. The normalization of the UHECR proton spectra is set as in the gamma-ray-induced cases.

A clear difference between gamma-ray and UHECR-induced cascades is the continuous injection of electrons through Bethe-Heitler pair creation during UHECR propagation in the latter case. UHECRs with $E_p\sim 10^{19}$ eV have energy-loss paths $\sim 1$ Gpc for Bethe-Heitler pair production, whereas 10 -- 100 TeV gamma rays only travel $\sim 3$ -- 200 Mpc before being absorbed by $\gamma\gamma$ production. This allows UHE electrons to be injected and cascade to such energies far from the source. Thus, some VHE photons can reach observers before being attenuated even when $\tau_{\gamma\gamma}(E,z)\gg 1$, resulting in a harder spectrum than expected for a gamma-ray-induced cascade. 

\section{Spectral Modeling and Scenario Distinction} \label{sec:spectra}

\begin{figure}
\includegraphics[width=1.\linewidth]{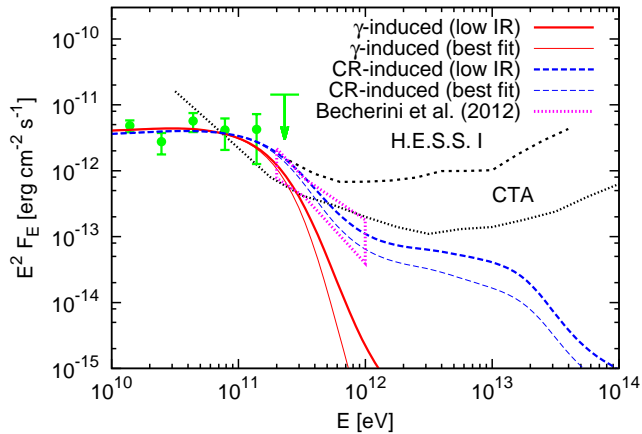}
\caption{SEDs calculated for gamma-ray-induced ({\it red}) and UHECR-induced ({\it blue}) cascade scenarios for KUV 00311$-$1938 ($z = 0.61$) using {\it low IR} ({\it thick}) and {\it best fit} ({\it thin}) EBL models deduced by \citet{Kneiske2004AA413p807} with the analyzed LAT data ({\it green}) with a H.E.S.S. preliminary spectrum \citep[{\it magenta};][]{Becherini2012AIP1505p490}. The isotropic equivalent energy of input gamma rays for the gamma-ray-induced cascade $L_{\gamma}^{\rm iso}$, and of UHECR source protons for a UHECR-induced cascade $L_p^{\rm iso}$ are $3.5 \times 10^{46}$ erg s$^{-1}$ and $1.1 \times 10^{47}$ erg s$^{-1}$, respectively. The differential sensitivity curve for a 50-hour observation with H.E.S.S. I (http://www.mpi-hd.mpg.de/hfm/HESS/pages/home/proposals/; {\it dashed line}), and the 50-hour sensitivity goal of the Cherenkov Telescope Array \citep[CTA;][{\it dotted line}]{Actis2011ExA32p193} are also plotted. The flux lower than the sensitivity curve can be achieved under a relaxed criterion of wider energy-bins and lower significance required to estimate flux in each bin.}
\label{fig:kuv00311sed}
\end{figure}

Figure \ref{fig:kuv00311sed} shows the SED of the BL Lac object KUV 00311$-$1938, the nearest object in the source list. This is the only source detected by IACTs in the \citet{Neronov2012arXiv1207.1962} sample at present. The hard spectrum obtained by LAT is well reproduced by both gamma-ray and UHECR-induced cascade scenarios between $10$ and 100 GeV. The UHECR-induced cascade predicts larger flux above $200$ GeV and harder spectrum than the gamma-ray-induced scenario above $\sim 1$ TeV. Preliminary H.E.S.S. data support the hadronic interpretations. Note that the redshift of this object is uncertain (see Section \ref{sec:discussions}).

We confirmed that the SEDs of the other more distant sources in the list, excepting sources with steep spectra, namely PKS 0426-380 and PKS 2142-75, are reproduced by both gamma-ray-induced and UHECR-induced cascade scenarios for the quoted redshifts. More distant sources allow the possibility to distinguish the two scenarios clearly by the difference in predicted spectral fluxes above $\sim 1$ TeV.  Due to their large distances, a sharper cutoff of the gamma-ray-induced spectra compared to the UHECR-induced spectra is predicted at the characteristic EBL absorption energy $E_c$ \citep{Murase2012ApJ749p63}, and a plateau of emission extending to $> 10$ TeV is predicted in the hadronic scenario.

In general, differential sensitivity is defined more conservatively than integral sensitivity for IACTs. Conventionally the differential sensitivity requires a $5 \sigma$ signal for a 50-hour observation in each of four equal-width logarithmic bins per decade, whereas the integral sensitivity is defined as a $5 \sigma$ excess of gamma rays above a given threshold energy for a 50-hour observation \citep[e.g.,][]{Aleksic2012APh35p435}. Thus, integral flux is more sensitive to the scenario distinction.

\begin{figure}
\includegraphics[width=1.\linewidth]{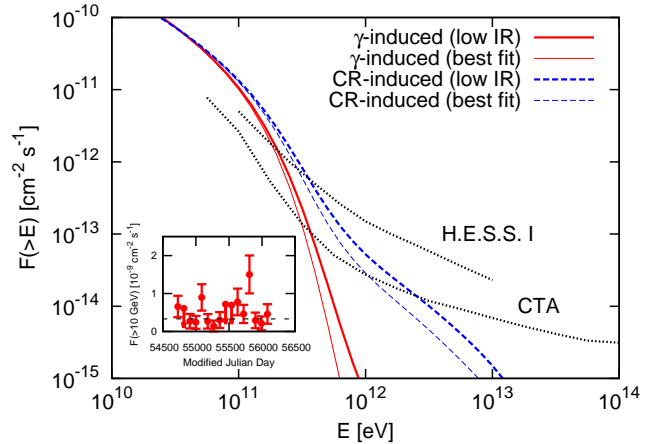}
\caption{Integral flux corresponding to the SEDs in Fig. \ref{fig:kuv00311sed} (KUV 00311$-$1938) with the H.E.S.S. I integral sensitivity (presented by Y.~Becherini in Rencontres de Moriond 2009; http://moriond.in2p3.fr/J09/) and the integral sensitivity goal of CTA for a 50-hour observation \citep{Actis2011ExA32p193}. The inset shows a $> 10$ GeV light curve with 16 equal time bins, each lasting 90.3 days. The light curve is consistent with a constant flux hypothesis with $\chi_r^2= 0.95$ which is calculated only from finite flux points.}
\label{fig:kuv00311int}
\end{figure}

Figure \ref{fig:kuv00311int} shows the integral flux corresponding to the predictions in Figure \ref{fig:kuv00311sed}. Here, we can obviously recognize that the UHECR-induced scenario can be distinguished from the gamma-ray-induced scenario by CTA. This source is detectable at the $5 \sigma$ level up to $\sim 3$ TeV for the {\it low-IR} model and $\sim 1$ TeV for the {\it best-fit} model in the UHECR-induced scenario, while it should only be detected up to $\sim 500$ GeV in the gamma-ray-induced scenario. Detection of this source above 1 TeV would be very strong evidence for hadronic origin of the radiation.

\begin{figure}
\includegraphics[width=1.\linewidth]{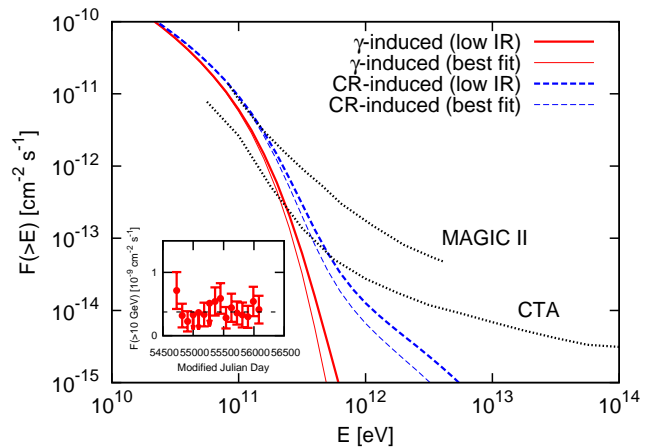}
\caption{Same as figure \ref{fig:kuv00311int}, but for PG 1246+586 ($z = 0.847$). $L_{\gamma}^{\rm iso} = 7.5 \times 10^{46}$ erg s$^{-1}$ and $L_p^{\rm iso} = 2.0 \times 10^{47}$ erg s$^{-1}$. The inset is a light curve similar to Fig. \ref{fig:kuv00311int}, with  $\chi_r^2= 0.40$ for a constant flux hypothesis.}
\label{fig:pg1246int}
\end{figure}

We demonstrate this behavior for a more distant source, PG 1246+586, in Figure \ref{fig:pg1246int}. Despite its distance, this source can be detected by CTA  below $\sim 200$ GeV for both scenarios. It is possible to distinguish between the two scenarios because the difference in detecting photons for the two scenarios would be larger than the range of uncertainties implied by the EBL models used, even with the flux of the characteristic hadronic plateau at high energies being below the CTA sensitivity. Thus, even gamma-ray sources with $z \sim 0.85$ can be utilized to disentangle the two scenarios. Other sources detectable with 50-hour observations with CTA in the source list are Ton 116, B3 1307+433, 4C +55.17, and PKS 1958-179. Note that the sensitivity of CTA North may be somewhat worse above $\sim 10$ TeV because no small-size telescopes are projected to be a part of the array.

\section{Discussion} \label{sec:discussions}

An important uncertainty for this study is the source redshift. It is  difficult to determine accurate photometric redshifts of BL Lac objects  owing to the strong nonthermal emission from relativistically beamed jets concealing the host-galaxy or broad-line region emission lines  \citep[e.g.,][]{Rau2012A&A538A26}. For instance, a recent detailed spectroscopic study of KUV 00311$-$1938 identifies no line related to the host galaxy, and provides a conservative lower limit of $z \geq 0.506$ from a Mg II doublet of narrow lines in intergalactic medium \citep{Pita2012arXiv1208.1785}, contrary to the often quoted value of $z = 0.61$ \citep[but large uncertainty is mentioned in][]{Piranomonte2007A&A470p787}. See also \citet{Shaw2013ApJ764p135}.

Uncertainty in redshift measurement results in two major problems for our study. The first is the resultant uncertainty of the predicted VHE fluxes of a source. A lower limit on redshift still remains useful because the UHECR-induced cascade scenario could be favored if VHE gamma rays are detected above energies where the flux in the gamma-ray-induced scenario is already suppressed assuming the lower-limit redshift. In other cases, a well-determined source redshift may be required to disentangle the two scenarios, as scenario discrimination depends both on the shape of the spectral cutoff and the cutoff energy $E_c$, which depends on the source redshift. The second problem is the resulting uncertainty in the total luminosity of gamma rays and/or protons to reproduce the observed flux. Recently, \citet{Padovani2012MNRAS422L48} pointed out that several distant HBLs have isotropic equivalent gamma-ray luminosities exceeding the trend of the blazar sequence \citep[e.g.,][]{Fossati1998MNRAS299p433,Kubo1998ApJ504p693} based on photometric redshift measurements. The same is true of some blazars in the list in the gamma-ray-induced scenario if their quoted redshifts are correct. Accurate redshift of sources is therefore also important to check the blazar sequence.

Although we have focused on spectral information to discriminate between the two possibilities, variability is also another critical clue. Cascade components cannot have short variability timescales, since the shortest time scales are  $\sim 1~{\rm yr}~{(E_\gamma/10~{\rm GeV})}^{-2} B_{-18}^{2}$ in the gamma-ray-induced case and  $\sim 10~{\rm yr}~{(E_\gamma/10~{\rm GeV})}^{-2} B_{-18}^{2}$ in the UHECR-induced case, respectively  \citep{Murase2012ApJ749p63}.  Note that, in the gamma-ray-induced case, cascaded gamma rays should be regarded as a mixture of attenuated and cascade components. Therefore, strong variability is possible if the cascade component is suppressed, whereas not in the UHECR-induced case. In our sample, there is no strong evidence for fast variability in KUV 00311$-$1938 and PG 1246+586 above $10$ GeV (see Figs. \ref{fig:kuv00311int} and \ref{fig:pg1246int}). Variability is weaker at the lowest photon energies of the cascade. By equating $z/H_0$ with the Thomson energy-loss timescale of relativistic electrons in the CMB, one finds that the cascade from blazars at low redshift ($z\ll 1$) extends to photon energies as low as $\approx 2.5$ keV$/(z/0.1)^2$.

One potential drawback of the UHECR-induced cascade scenario is that it may require large powers emitted in UHECRs. This is especially true if UHECRs are not beamed. If the blazars are embedded in magnetized structures, i.e., filaments and clusters of galaxies, as naturally expected, the jet-corrected CR luminosity required to reproduce the observed flux is about one and two orders of magnitude larger than in the case of no surrounding magnetic field, respectively, because of the isotropization of CR protons in the magnetic structures \citep{Murase2012ApJ749p63}. \citet{Razzaque2012ApJ745p196} derive lower limits of UHECR power following this scenario for extreme HBLs with known redshift, and point out that the required power is comparable with or in some cases exceeds the Eddington luminosity of the supermassive black hole (SMBH) of the blazars and if UHECRs are isotropized. Some blazars considered in this study may require UHECR powers close to the Eddington luminosity of SMBH mass $10^9 {\rm M}_{\odot}$.

We have considered IGMFs weak enough so as not to affect the spectrum above $10$ GeV. For stronger IGMFs, the cascade radiation in both the hadronic and leptonic scenarios becomes harder due to fewer low-energy photons made in the jet direction. This hardening may, however, be hidden because of intrinsic source gamma rays with energies $E \lesssim E_c$. Given a sharp spectral cutoff in the gamma-ray-induced scenario, it is clear evidence of the hadronic scenario and therefore the sources of UHECRs that a spectrum harder than that predicted from the gamma-ray-induced scenario is observed.

Finally, we mention that signatures of UHECR-induced and gamma-ray-induced cascades in the diffuse gamma-ray background are also distinguishable \citep{Murase2012JCAP08p030}.

To summarize, the SEDs of many of the distant blazars listed in \citet{Neronov2012arXiv1207.1962} can be modeled by both gamma-ray-induced and UHECR-induced cascade scenarios, as shown here explicitly for KUV 00311$-$1938 and PG 1246+586. Some of these blazars are predicted to be detectable by CTA, and their spectral properties will permit these two scenarios to be distinguished by the different detection threshold energies and spectral behaviors. Preliminary H.E.S.S. observations favor the hadronic interpretation for KUV 00311$-$1938 if at redshift $z = 0.61$. 

\acknowledgements 
We thank M.~Hayashida, K.~Kotera, and T.~Saito for discussions, M. Ajello and A.~Reimer for questions, and B.~Lott for a careful reading of the paper. We would also like to thank S.~Digel for corrections and constructive comments. H.~T.\ is supported by Japan Society for Promotion of Science. The work of C.~D.~D.\ is supported by the Office of Naval Research. 

The \emph{Fermi} LAT Collaboration acknowledges support from a number of agencies and institutes for both development and the operation of the LAT as well as scientific data analysis. These include NASA and DOE in the United States, CEA/Irfu and IN2P3/CNRS in France, ASI and INFN in Italy, MEXT, KEK, and JAXA in Japan, and the K.~A.~Wallenberg Foundation, the Swedish Research Council and the National Space Board in Sweden. Additional support from INAF in Italy and CNES in France for science analysis during the operations phase is also gratefully acknowledged.

\bibliographystyle{apj.bst}

\end{document}